# Quasi-Phase-Matching Enabled by van der Waals Stacking

Yilin Tang[1,2#], Kabilan Sripathy[3,4#], Hao Qin[1#], Zhuoyuan Lu[1], Giovanni Guccione[2,3], Jiri Janousek[1,2], Yi Zhu[5], Md Mehedi Hasan[1], Yoshihiro Iwasa[6], Ping Koy Lam[3,7*] and Yuerui Lu[1,2*]

[1]School of Engineering, College of Engineering, Computing and Cybernetics, the Australian National University, Canberra, ACT, 2601, Australia

[2]ARC Centre for Quantum Computation and Communication Technology, the Australian National University, Canberra, ACT, 2601, Australia

[3]Research School of Physics, College of Science, the Australian National University, Canberra, ACT, 2601, Australia

[4]Department of Computer Engineering, TUM School of Computation, Information and Technology, Technical University of Munich, 80333 Munich, Germany

[5]Clarendon Laboratory, University of Oxford, Oxford, OX1 3PU, United Kingdom

[6]RIKEN Center for Emergent Matter Science, Hirosawa 2-1, Wako, Saitama 351-0198, Japan

[7]Quantum Innovation Centre (Q.InC), Agency for Science Technology and Research (A*STAR), 2 Fusionopolis Way, Innovis #08-03, Singapore 138634, Republic of Singapore

# These authors contributed equally to this work.
* To whom correspondence should be addressed: Ping Koy Lam (Ping.Lam@anu.edu.au), Yuerui Lu (yuerui.lu@anu.edu.au)

**Abstract:** Quasi-phase matching (QPM) is a technique extensively utilized in nonlinear optics for enhancing the efficiency and stability of frequency conversion processes. However, the conventional QPM relies on periodically poled ferroelectric crystals, which are limited in availability. The 3R phase of molybdenum disulfide (3R-$MoS_2$), a transition metal dichalcogenide (TMDc) with the broken inversion symmetry, stands out as a promising candidate for QPM, enabling efficient nonlinear process. Here, we experimentally demonstrate the QPM at nanoscale, utilizing van der Waals stacking of 3R-$MoS_2$ layers with specific orientation to realize second harmonic generation (SHG) enhancement beyond the non QPM limit. We have also demonstrated enhanced

**spontaneous parametric down-conversion (SPDC) via QPM of 3R-MoS$_2$ homo-structure, enabling more efficient generation of entangled photon pairs. The tunable capacity of 3R-MoS$_2$ van der Waals stacking provides a platform for tuning phase-matching condition. This technique opens interesting possibilities for potential applications in nonlinear process and quantum technology.**

## Introduction

The approach of QPM was first proposed in the 1970s as a theoretical concept and bloomed in the 1980s with the advent of periodic poling of nonlinear crystals. The conventional method applies a periodically varying electric field into ferroelectric crystals,[1] such as lithium niobate (LiNbO$_3$), lithium tantalate (LiTaO$_3$), and potassium titanyl phosphate (KTiOPO$_4$), which have been fabricated with a periodic grating structure.[2-4] This is accomplished through the use of microstructural electrodes that apply a strong electric field to the crystal for a specified duration, resulting in a permanent reversal of crystal orientation and the opposite nonlinear coefficient's polarity solely beneath the electrode fingers.[5] The periodic reversal occurs at coherence length within the crystal, where the phase mismatch between the interacting fields would lead to periodic constructive and destructive interference. Upon reversal of the domain, the effects of the phase mismatch are rewound and a positively constructive nonlinear interaction is restored.[6,7] There are a few long-standing challenges for traditional QPM techniques. Firstly, traditional QPM relies on ferroelectric nonlinear crystal materials, which introduces limitations in terms of operating conditions due to limited choice of ferroelectric crystals and their strong optical dispersion and birefringence.[1] Secondly, ferroelectric crystals are typically sensitive to environmental factors such as temperature, humidity, or mechanical stress, which gradually degrade the domain structure and eventually reduce the efficiency of the nonlinear crystal over time.[8] Thirdly, the bulkiness of ferroelectric crystals and challenges in integration with photonic chips, along with the complex fabrication process requiring the precise poling and grating formation, raise cost and complexity of the batch fabrication.

Ongoing efforts aim to address these challenges through new materials and fabrication techniques for integrated QPM devices. Two dimensional (2D) layered materials stand out as a promising candidate for QPM due to their thinness and ease of integration. These ultra-thin crystals exhibit a favourite SHG to weight ratio[9-11] and giant optical nonlinearity[12-15] which can be orders of magnitude stronger than that of equivalent bulk materials,[16,17] and offer a range of advantages over traditional ferroelectric materials. Two of the most common crystal structures observed in transition metal dichalcogenides (TMDc) are the 2H and 3R phase[18-20]. The nonlinearity of these 2H TMDc stacks is nullified for even number layers, while odd-numbered layers retain some nonlinearity, which is generally weaker than that of a single monolayer. A different stacking structure is found in the 3R polytype, which exhibits a periodic translational symmetry that repeats every three layers while maintaining a consistent orientation across those layers (Fig. 1a). This polytype has been successfully used to enhance the total nonlinear gain of the TMDc nano-crystals by preserving the broken inversion symmetry and, therefore, constructively building up the atomic nonlinear dipole with increasing layer thickness.[18-20] Thus, unlike the 2H phase where the SHG signal does not scale with the number of layers, for the 3R phase, the intensity of SHG signal shows a quadratic relation, as would be expected for a phase-matched interaction.[21] Although quadratic at the outset, the enhancement slowly reverses trend as the number of layers grows and the phase mismatch increases, until a coherence length is reached where the SHG signal reaches a plateau due to reaching the non QPM limit.[22]

In this paper, we demonstrate nanoscale QPM on 2D TMDc crystals by superimposing 3R polytype samples into particular QPM stacking order reminiscent of periodic poling. Twisting the relative orientation of the stacks, we demonstrate a significant enhancement in conversion efficiency beyond regular stacks of similar combined thickness. Our subsequent endeavors of its reverse process SPDC by employing QPM approach demonstrated the efficient generation of entangled photon pairs in TMDs. This technique of using van der Waals stacking of 3R-$MoS_2$ enables fine-tuning of the phase-matching condition and provides a larger interaction length to realize efficient and tunable nonlinear optical processes, with potential applications in areas such as frequency conversion,[23] optical switching, and quantum communications.[24-26] This research direction offers an alternative to QPM with periodically poled crystals and opens up promising applications based on the second harmonic (SH) generation and entangled photon pairs generation in 2D materials. It also identifies the need to find suitable 2D materials for nonlinear optics with stability, controllable layer numbers, and strong second order nonlinear coefficients for future high-performance optoelectronic and integrated devices.

## Results

**Modelling of QPM by van der Waals stacking**

Figure 1 shows the theoretical model of a unique QPM enabled by van der Waals stacking, using 3R-$MoS_2$ as an example. The SHG signal of 3R-$MoS_2$ increases with the increase of its thickness until the coherence length ($t_c$) is reached, where the phase mismatch is caused by the difference in refractive index for the seed and the second harmonic fields (Fig. 1a). In the limit of negligible absorption of the fundamental, from the model used in the referenced study the SHG signal with phase mismatch model can be expressed as:[22,27]

$$I_{2\omega}(t) = \frac{2\omega^2 d_{eff}^2 t^2}{\epsilon_0 n_{2\omega} n_\omega^2 c^3} I_\omega^2(0) Sinc^2(\Delta k t/2) \quad (1)$$

where $\omega$ is the frequency of fundamental pump, $\chi^{(2)}$ is the second order susceptibility term,

$d_{\text{eff}} = 2\chi^{(2)}$, $t$ is the sample thickness, $n_\omega$ and $n_{2\omega}$ are the refractive indices of 3R-MoS$_2$ at the fundamental wavelength (FW) field and its SH, respectively. The difference in refractive index induces a wavevector mismatch of $\Delta \mathbf{k} = \mathbf{k}_{2\omega} - 2\mathbf{k}_\omega$, where $\mathbf{k}_{2\omega}$ and $\mathbf{k}_\omega$ are the wavevectors of SH wave and FW, respectively. As the propagation length increases, the locally generated SH becomes progressively out of phase. The total SH intensity increases until the wave reaches a thickness equal to the coherence length ($t_c = \pi/\Delta \mathbf{k}$), where the phase mismatch equals to one pi and after which the converted field has opposite phase, therefore destructively interfering with waves from previous planes and reducing the SH intensity.

By stacking two van der Waals flakes A and B with a twist angle, the SHG signal with QPM model can be expressed as (Supplementary Note 1):

$$I_{qpm}(t_1, t_2) \propto (|(e^{i\Delta \mathbf{k} t_1} - 1) + \cos[3\theta] e^{i\Delta \mathbf{k} t_1}(e^{i\Delta \mathbf{k} t_2} - 1)|)^2 \qquad (2)$$

where $t_1$ and $t_2$ are the thicknesses of A and B, respectively, and $\theta$ is the twist angle between A and B (Fig. 1b).

In accordance with theory of QPM, enhancement is maximized when using a stack AB of two identical 3R-MoS$_2$ layers with thickness $t_c$. When the phases of the fundamental and SH waves are perfectly matched at coherence length $t_c$,[28,29] a constructive interference of the SH field is restored and the SH field continues to grow throughout the second 3R-MoS$_2$ layer. Considering the hexagonal structure and three-fold symmetry of 3R-MoS$_2$, the interference of SH waves generated from an AB stack shows a three-fold rotational symmetry as a function of $\theta$. When $\theta$ is equal to 60° (or, equivalently, 180º or 240º) the stack exhibits fourfold enhancement of SHG intensity when $\theta$ is equal to 60° (Supplementary Figure 1a). Our proposed approach involves stacking multiple layers of 3R-MoS$_2$ with thickness of $t_c$, with each layer oriented at an angle of 60° to the previous layer (Fig. 1c), creating a periodic modulation of the nonlinear susceptibility along the growth direction, facilitating QPM of the interacting waves.

**QPM using two layers of 3R-MoS$_2$ with tunable twist angle $\theta$**

In experiments, mechanically exfoliated 3R-MoS$_2$ samples (see Methods) were used to demonstrate the QPM by van der Waals stacking. Supplementary Figure 1b shows the measured SHG intensity from a specific 3R-MoS$_2$ flake (thickness 396 nm) as a function of the pumping power. The extracted power law index of the linear fitting function is 1.92, consistent with the expected quadratic power dependence of the SHG emission.[22] The transmitted SHG through a 3R-MoS$_2$ flake is highly dependent on sample thickness and both interference and phase-matching effects must be taken into account.[22] Owning to the finite reflectivity at the boundary interfaces, the transmissivity of the FW light is modulated by a Fabry-Pérot-like interference and varies periodically with the thickness of the 3R-MoS$_2$ (Supplementary Figure 1c black curve), leading to a pattern of peaks at regular intervals (Supplementary Note 2). These periodical peaks are modulated by the QPM envelope (Supplementary Figure 1c red curve). In order to realize the QPM by van der Waals stacking with maximized SHG intensity, the pump wavelength was set to 1550 nm (0.80 eV) yielding SHG centred at 775 nm (1.60 eV), which can enable a constructive interference for samples with thickness of $t_c$, or multiples of $t_c$ after stacking. At the same time, the energies of SHG and pumping photons are below the optical gap of MoS$_2$, avoiding absorption losses. In addition, 1550 nm is a significantly used wavelength in communication applications and hence is of interest. The measured SHG intensity was strongly dependent on the sample thickness (Supplementary Figure 1d), which was well fitted by simulation model considering the contributions from both inference and phase-matching effects (Supplementary Note 3). The coherence length $t_c$ for 3R-MoS$_2$ for 1550 nm to 775 nm (and vice versa) conversion was extracted to be 572 ± 5 nm, where the SHG conversion efficiency $\eta$ is extracted to be

$4.64 \times 10^{-6}$ (under power density 30 GW/cm² (Supplementary Note 4). The estimation based on that predicts that near-unity SHG efficiency can be achieved with ~494 3R-MoS$_2$ flakes (280 μm thickness), offering a potentially more compact alternative to PPLN crystals requiring device lengths of 2-3 cm (Supplementary Note 5).[30] Our experiments identified a damage threshold power density of 34.82 GW/cm² at 1550 nm, corresponding to damaging threshold power of 35 mW (Supplementary Note 6). This value is comparable to the previously reported damage threshold (around 30 GW/cm²) for multi-layer TMD samples.[22] The poling condition of 3R-MoS$_2$ van der Waals stacks is determined by the desired phase-matching wavelength, with coherence length $t_c$ calculated based on corresponding refractive indices (Supplementary Figure 4c, Supplementary Note 7). Our analysis shows a phase-matching bandwidth of approximately 640 nm at the pump wavelength 1550 nm (Supplementary Note 8). Employing a longer pump wavelength could potentially broaden this bandwidth by shifting the SHG signal away from the exciton resonance of 3R-MoS$_2$ (around 680 nm).

Then we tested the QPM as a function of twist angle $\theta$, by using two 3R-MoS$_2$ samples, A and B, of thickness ~$t_c$ placed onto sapphire substrates. The position of the bottom sample A (facing up – where the sample flake sitting on top of the sapphire substrate and the beam transmits through the flake then the sapphire substrate) was fixed, and the top sample B (facing down- the beam transmits through the sapphire substrate then the flake) was positioned and rotated using a nano precision stage, by keeping a constant air gap of ~10 μm between A and B during the rotation (Fig. 2a). At each twist angle (10° intervals), the transmitted SHG intensities were measured from three areas, AB, A and B (Fig. 2b). Since A is stationary, its SHG is not expected to vary as a function of twist angle and its SHG pattern hence appears as a circle. The SHG signal of B alone exhibits a sinusoidal pattern with six-petal polar plot with lobes corresponding to the increased SHG efficiency when the armchair axes of the crystals align with the FW field polarization, and SHG suppression when the zigzag directions align with polarization instead.[31] The SHG from stacked area AB showed clear three-fold symmetry (Fig. 2b), indicating the occurrence of QPM with peaks when the crystal lattices of the two samples are aligned ($\theta$ = 60°) and troughs occur when they are anti-aligned (0°), consistent with theoretical predication (Supplementary Figure 1a). We also used the transfer matrix method to fit the measured SH intensity from area AB (Fig. 2b and Supplementary Note 9) and the air gap was extracted to be 9.72 μm, consistent with our experimental setting. Simulations of the airgap length (Supplementary Figure 4a) confirm minimal influence of cavity effect, further supporting the observed enhancement (Supplementary Note 10).

**Experimental demonstration of QPM via 3R-MoS$_2$ stacking**

To further confirm QPM enabled by van der Waals stacking, two 3R-MoS$_2$ flakes A and B (with the thickness close to $t_c$) were physically stacked together without airgap, with a twist angle $\theta$ of 60° (Fig. 3a). The thickness of A, B and AB were measured to be 576, 564 and 1140 nm, respectively (Fig. 3a inset). Angle resolved SHG was also performed to precisely determine the crystal orientation and the twist angle between the two flakes.[32] The measured polar plots in Figure 3b confirm that the lattice of the A and B are aligned at a twist angle of 60°, at which a maximum SHG was observed from the AB homo-structure. The maximum SHG signal from AB was measured to be 3.8 times higher than that from the two individual flakes A and B, consistent with the theoretical prediction - the final SHG intensity increases quadratically with the number ($N$) of stacks (SHG $\propto N^2$) (Supplementary Figure 2e, Supplementary Note 1). For optimal QPM in two-layer 3R-MoS$_2$ stack, flake thicknesses should ideally be close to the coherence length, with a tolerance range (Supplementary Figure 4b, Supplementary Note 10). Deviations from this range can significantly reduce the efficiency of SHG. More importantly, the recovery of SHG at the total combined thickness of the two 3R-MoS$_2$ flakes is a clear indication of the QPM condition being satisfied where, without intervention, zero SHG signal would otherwise be observed. Similarly, a stack with a non-optimal twist angle of 45° was also fabricated and characterized for comparison (Supplementary Figure 2). Angle resolved SHG experiments were also conducted to confirm the 45° stacked angle (Supplementary Figure 2b). The measured maximum SHG intensity from 45°-twisted-AB stack was 2.7 times of SHG from an individual flake with thickness close to $t_c$ (Supplementary Figure 2b), also consistent with theoretical prediction (Supplementary Figure 2e). Supplementary Figure 2e shows the comparison summary of the measured and simulated SHG intensities of the A, B and AB areas, from three 3R-MoS$_2$ homo-structures with twisted angles of 60°, 45° and 0°. This twisted homo-structure by van der Waals stacking offers a unique approach of periodic poling, enabling the high performance of nonlinear optoelectronic devices. Our angle resolved SHG measurement (Fig. 3b) on thick 3R MoS$_2$ did not show the previously reported observation of two longer petals than the other four. We attributed the difference to be mainly from the sample thickness. Our flakes have the thickness close to $t_c$ (~572 nm), significantly thicker than those used in previous studies.[19,22] Consequently, the unequal SHG response observed in thinner flakes could be attenuated in our thicker samples. Any remaining asymmetry due to staggered stacking might contribute a small enhancement to two of the petals, which could potentially fall within the range of system error (around 5%).

To further realize the integrated devices, we have successfully demonstrated a three-layered stack of 3R-MoS$_2$ (each layer with thickness close to $t_c$) through a meticulous design with quadratic enhancement of SHG by means of constructive interference of the SH waves (Supplementary Figure 4d inset). Layer 1 and layer 2 were stacked together to form a homo-structure with a twist angle of 60°, as shown in Figure 3a. Layer 3 has a measured thickness of 570 nm, close to the coherence length $t_c$. Layer 3 could be rotated with a tunable angle $\theta_3$, as there is a physical gap of ~9.30 μm between layer 3 and layer 2 in experiment. The measured SHG from three-layer-stacked area showed clear three-fold symmetry (Supplementary Figure 4d), indicating the occurrence of QPM constructive interference when the lattices of the third sample align with the bottom layer ($\theta_3 = 0°$) and destructive when they are anti-aligned (60°), which can be well fitted by our QPM modelling (Supplementary Figure 4d, Supplementary Note 11). Comparing with the SHG from a single layer with a thickness of length $t_c$, the measured maximum SHG intensity from the three-layer-stacked area showed an enhancement factor of 7.5, which is slightly lower than the expected enhancement factor of 9 when three layers are physically stacked without any airgap as indicated by Figure 1c. This discrepancy could be attributed to the presence of airgap and the divergence of collimation path. Nonetheless, an enhancement of 7.5 relative to a single layer is an outstanding indication of the recovery of quadratic enhancement in repeated QPM stacks and the preservation of coherent superposition of the SH waves generated in the different layers, substantially improving the overall efficiency of the SHG process. To further validate the established model and eliminate the influence of airgap, we stacked a three-stack 3R-MoS$_2$ homostructure (Supplementary Figure 4e). Experimental measurements revealed a roughly ninefold enhancement in the combined SH signal from the three-layer homostructure region compared to the single flake. This finding aligns well with the theoretical calculations SHG ∝ $N^2$ (Fig.3c). In recent research by engineers exploring robotic arm manipulation for large-scale periodic poling of TMDs offers a promising future for van der Waals stacks, enabling precise control over layer sequence, twist angle, and large-scale, automated fabrication of complex devices with tailored properties (Supplementary Note 12).[33] This could lead to significant miniaturization of SHG crystals, which would be useful for microchip integration and space applications.

**Enhanced SPDC via QPM of 3R-MoS₂ homo-structure**

Following the successful demonstration of SHG enhanced through QPM, we subsequently conducted an experiment of its reverse process, SPDC. SPDC is solely a quantum process where a single photon $\gamma_p$ with higher energy undergoes a nonlinear interaction within the crystal, yielding the entangled photon pairs, known as the signal photon $\gamma_s$ and idler photon $\gamma_i$, in accordance with the conservation of energy and momentum (Fig. 4a). Recently, SPDC was observed from low-dimensional thin materials,[25,26,34,35] however, their conversion efficiencies are relatively low compared with conventional periodically poled structures with QPM.[36,37] Therefore, van der Waals QPM could be a potential way to increase their efficiencies. The SPDC process was observed on the 3R-MoS₂ homo-structure with 60° stacked angle with respect to the preceding one, as depicted in Figure 3a, using a continuous-wave laser at 775 nm (Fig. 4b). Detailed description of the setup can be found in Methods. We tested the photon pair correlations via the Hanbury, Brown and Twiss interferometry by employing a 50/50 multi-mode fibre coupler, whose outputs were connected to a pair of single photon avalanche diodes (SPADs), and photon arrivals were recorded using a Time-Correlated Single Photon Counting (TCSPC) system. Figure 4c presents the second-order correlation function $g^{(2)}(\tau)$ measured from the 3R-MoS₂ stacked area AB, individual 3R-MoS₂ area A and B, and a sapphire substrate, where a prominent peak at zero delay time was only observed for the 3R-MoS₂ samples.[38] The two-photon correlation peak has a peak-to-background ratio well above 2 at zero time delay ($g^{(2)}(0)$ ~ 4.5 from 3R-MoS₂ stacked area AB, ~3.2 and ~2.8 from individual 3R-MoS₂ area A and B), providing a definitive indication of SPDC signature of correlated photon pair generation.[25,39,40]

Pump power dependent measurements were further conducted on A, B and AB areas. The linear relation was observed revealing the underlying nature of SPDC[25] and the extracted slope of AB is approximately fourfold higher than those from individual A and B (Fig. 4d). Considering the

QPM scheme, there is a similar trend of photon pair generation rate of stack AB as compared to those on individual flake A and B, which mirrored the observed SHG intensity enhancement. This is consistent with the theoretical model that SPDC efficiency proportionally correlates to the SHG intensity (Supplementary Note 13). As a consequence, the photon pairs generation can be phase matched at coherence length $t_c$ (Supplementary Note 13) and we experimentally demonstrated that the SPDC process can be engineered through QPM. As the number of flakes increases, the linewidth of the resonance narrows, consistent with the findings of Trovatello et al.[41] Our study further extends these observations, and the bandwidth of SPDC is estimated to be approximately 289 nm, 189 nm and 138 nm with the flake thickness at $t_c$, $2t_c$ and $3t_c$ respectively (Supplementary Note 8).

**Discussion**

We demonstrated a promising way to generate QPM by van der Waals stacking. Experimental results highlight the potential of fabricating an incredibly small and lightweight van der Waals nonlinear crystal with QPM and a superior SHG to weight ratio, surpassing their bulky 3D counterparts.[32] Through both contact and non-contact measurements at 1550 nm, along with the subsequent angle resolved orientation measurements, QPM between the two layers of 3R-MoS$_2$ (both thickness close to coherence length $t_c$) has been demonstrated, realizing QPM and strongly enhanced SHG greatly beyond the non QPM limit. These demonstrations exhibit strong agreement with the derived angle resolved QPM model. Additionally, the homo-structures with a tunable twist angle SHG responses correlates well with the periodically poled QPM model, making 3R-MoS$_2$ as a remarkable candidate for 2D material SHG devices, such as a two-layer SHG crystal with tunable phase-matching condition generated by van der Waals stacking with a tuanble twist angle. We have also demonstrated enhanced SPDC via QPM of 3R-MoS$_2$ homo-structure. This accomplishment further corroborates the successful improvement of nonlinearity via QPM techniques. The periodic modulation of the optical properties in the van der Waals stacking provides an innovative and general way to realize QPM without the requirement of any ferroelectric crystals. The adjustability of the twist angle in van der Waals stacking QPM holds significant promise in engineering SPDC-based quantum devices aimed at producing tailored quantum states like Bell and N00N states for applications in quantum optics and communication.[42] To give an example, crafting such a device would necessitate the precise configuration of two sets of nonlinear material, each with a specific thickness, and arranged such that their crystal axes are perpendicular to one another – orthogonal quasi-phase matching (OQPM).[43] It is very challenging to use conventional periodic poling to realize OQPM, but it would be much easier for van der Waals stacking QPM to create OQPM with twist angles of (0°, 30°, 60°, 90°,0°, 30°, 60°, 90°…). It can enable efficient energy transfer and enhances nonlinear optical properties, thereby making these 2D crystals a promising platform for nonlinear optical applications such as a typical frequency doubling crystal, interferometric auto correlators, biological indicators,[44] the generation of squeezed states,[45] quantum information processing[24,25] and a whole host of other applications.

**Methods**

**Sample fabrication.** Bulk 3R-MoS$_2$ crystal was grown using chemical-vapor transport technique.[18] The 3R-MoS$_2$ flakes were mechanically exfoliated onto PDMS gel film, and then dry transferred onto 400-μm-thick sapphire substrates. Sapphire substrates were preferred over traditional silicon substrates due to their transparency. Thickness was measured using atomic force microscopy (AFM) and phase shift interferometer (PSI) for thinner exfoliated flake (up to 10 nm), along with optical contrast. The thicker flakes (10 nm-2000 nm) were characterized using Vertical Scanning Interferometry (VSI) and the surface profiler. The thickness profile was acquired from high-precision data acquisition components to measure the thickness of the distinct regions through physical contact.

**Thickness characterization.** Data has been collected across almost 300 sample regions with thicknesses ranging from roughly 21 nm to 1444 nm, characterized by VSI and Surface profiler. Thin sample thicknesses were determined using AFM (Bruker) and processed via Nanoscope. The determination of thicker sample thicknesses was accomplished through employment of Surface Profiler in conjunction with VSI.

**Angle resolved phase matching measurement.** The custom-built angle resolved phase matching setup consists of a 100 fs tunable pulsed laser (Spectra-Physics Mai Tai) operating with an OPO to output the fundamental pump wavelength at 1550 nm, a broadband $\lambda/2$ waveplate, a polarizer, and a rotatable 6-DOF nanopositioning stage (SmarAct SmarPod) to align the two flakes for QPM. The $\lambda/2$ waveplate served to align the polarization of pump beam with the armchair axis of the bottom crystal which was held static on the translation stage. The samples were excited and measured via two identical ×50 near infrared (NIR) infinity corrected objective lenses (NA = 0.42), and the signal was collected in the transmission mode. The polarizer was put at the emission side to select the polarization component of the second harmonic (SH) radiation parallel to the polarization of the pump beam. The top crystal was mounted and positioned using the nano precision stage and the two samples were aligned to be parallel and brought within 9-11 μm of each other. The distance was determined by measuring the distance between their focus planes. The foci of the top and bottom objectives were positioned in between the two samples to obtain the highest SH intensity of the stacked region. The alignment process involved positioning the sapphire chips in parallel and focusing onto the overlapping stacked area to achieve maximum SH signal. The bottom objective lens was then aligned to optimize the SH signal. The focus was kept the same throughout the measurements. Detection of the SH signal was accomplished using a spectrometer (Andor Kymera-spectrograph and iDuS 416-CCD) with an 800 nm short pass filter to block the fundamental pump beam. The measurements were taken from three characterization areas, top sample, bottom sample and the stacked area while the top sample was rotated at 10° intervals. The signal from the bottom sample, which theoretically should remain constant was monitored to ensure consistency and to identify and rectify any misalignment. The angle resolved SHG was also conducted to confirm the lattice orientation. Both λ/2 waveplate and polarizer are mounted on electronic rotation mounts allowing synchronized positioning for copolarization.

**Van der Waals stacking device fabrication.** Before stacking, the top flake was on a polydimethylsiloxane (PDMS) soft substrate with the nano precision stage and the other bottom flake was on a sapphire rigid substrate. Prior to the transfer, the angle resolved QPM measurement was also conducted in order to the alignment of stacked orientation of the two flakes.

**Spontaneous parametric down-conversion.** To capture the two-photon correlation properties, a custom-built SPDC setup with two SPADs (PDM-IR, Micro Photon Devices) and one TCSPC with 1 ps resolution (HydraHarp 400, PicoQuant) was implemented. A 1550 nm continuous-wave laser (NKT Koheras ADJUSTIK) was sent to a fibre amplifier (NKT Koheras BOOSTIK). The 775 nm pump laser was generated through SHG from a periodically poled Potassium titanyl phosphate (PPKTP) crystal placed into a bow-tie cavity, and the pump field was filtered from residual fundamental using a 1000 nm short pass filter. After controlling the power of the pump beam through a polarizing beamsplitter and a 780 nm λ/2 waveplate, the pump was focused onto the sample by a B-coated Best Form spherical lens ($f = 75$ mm). Subsequently, a C-coated high NA aspheric lens ($f = 20$ mm, NA = 0.54) was used to collect the signals. The residual pump beam was initially filtered through two AR coated mirrors (99% transmission at 775 nm and 99% reflection at 1550 nm). Further filtering of the SPDC signal from the pump and photoluminescence photons was achieved using two 1550 nm bandpass filters with a full-width at half-maximum of 12 nm and two 1500 nm high pass filters. The SPDC signal was coupled into a $1 \times 2$ 50:50 multimode fibre optic coupler (Thorlabs, TM50R5F1B) using a mode-matching lens ($f$ = 40mm, AR at 1550 nm) and a fibre collimator (Thorlabs, F240FC-1550). The outputs of the fibre coupler were connected to two SPADs operating in free-running mode with 50 μs hold-off time. We employed a TCSPC system to record arrival of each photon onto the two SPADs and performed data post-analysis using 1 ns bin time window to reveal photon pair correlations.

Due to the nature of the single photon avalanche diodes (SPADs), a hold-off time must be configured in free-running mode to suppress after pulsing, which results in a lowered the duty cycle of the SPADs. To rule out the loss of efficiency caused by the hold-off time, we used the following formulas for coincidence count rate ($C_{correct}$ [s$^{-1}$]) and SPAD count rate ($N_{i_{correct}}$ [s$^{-1}$]) to correct for the detectors' deadtime:[46]

$$C_{correct} = \frac{C_{raw}}{(1-N_1\tau_1)(1-N_2\tau_2)}$$

where $C_{raw}$ is the measured coincidence count rate reading. $N_i$ and $\tau_i$ are the measurements of SPAD count rate and hold-off time respectively. In our experiment, the hold-off time was set to be 50 μs.

**Data availability**

Relevant data supporting the key findings of this study are available within the article and the Supplementary Information file. All raw data generated during the current study are available from the corresponding authors upon request.

**Acknowledgements**

The authors acknowledge funding support from ANU PhD student scholarship (Y.T., H.Q., Z.L., M. H.), Australian Research Council grant no. DP240101011(Y.L.), DP220102219 (Y.L.), DP180103238 (Y.L.), LE200100032 (Y.L.) and ARC Centre of Excellence in Quantum Computation and Communication Technology project number CE170100012 (Y.L., J. J.). The authors further acknowledge the funding support from JSPS grant no. 19H05602 (Y. I.).

**Author Contributions**

Y. L., P. L. and G. G. conceived and supervised the project; Y. T. and Z. L. prepared 3R-MoS$_2$ samples and homo-structure device; Y. T. and K. S. carried out the optical measurements; Y. T., K.S. and Y. L. analyzed the data; H.Q., K.S., G. G., Y. Z. and J. J. built up the SHG optical setup; H. Q. and J. J. built up the SPDC set up and took the SPDC measurements; M. H. took the AFM imaging; Y. I. provided 3R-MoS$_2$ crystal; Y. T., K. S., H. Q. and Y. L. drafted the manuscript and all authors contributed to the manuscript.

**Competing interests**

The authors declare no competing interests.

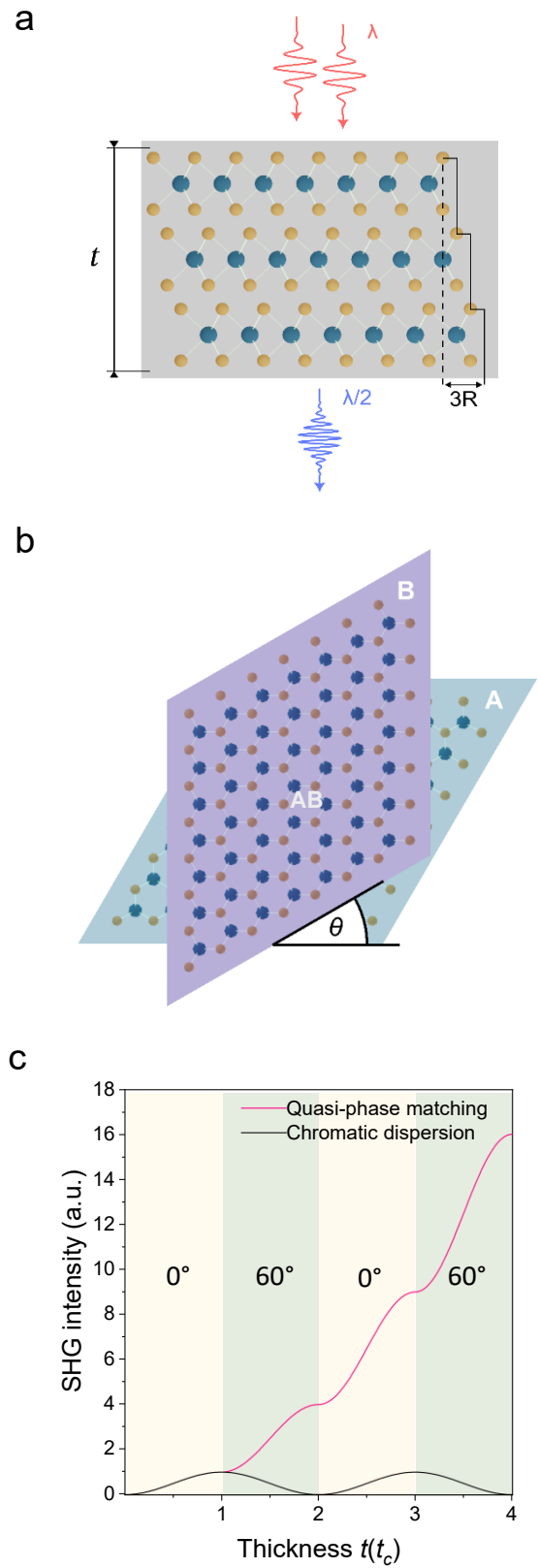

Figure 1

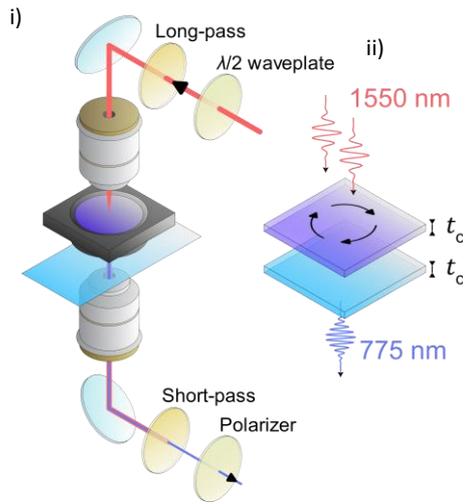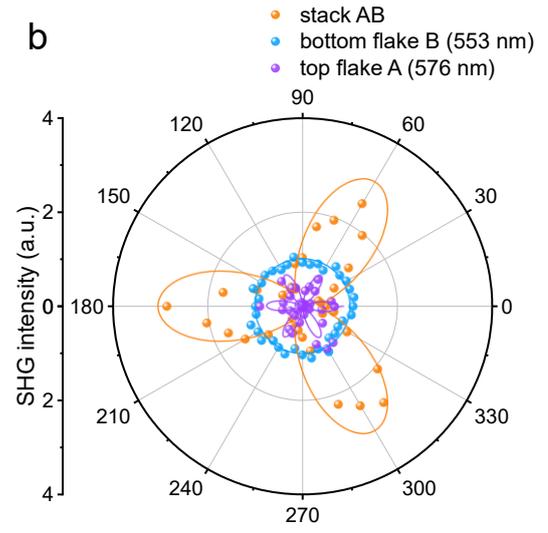

Figure 2

a

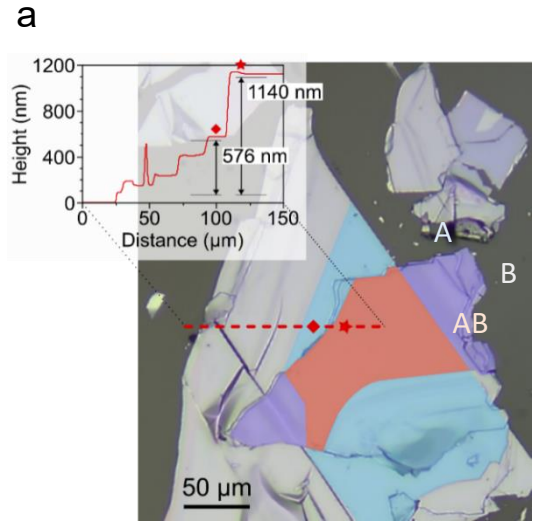

b

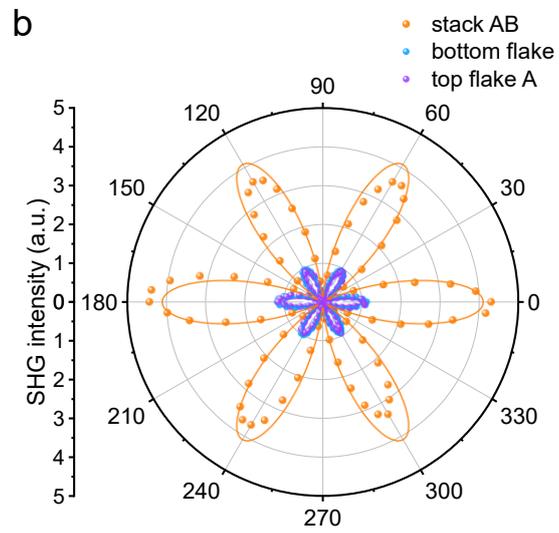

c

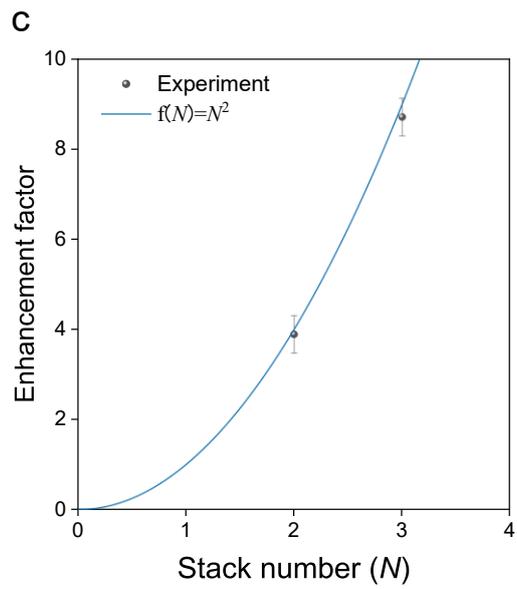

Figure 3

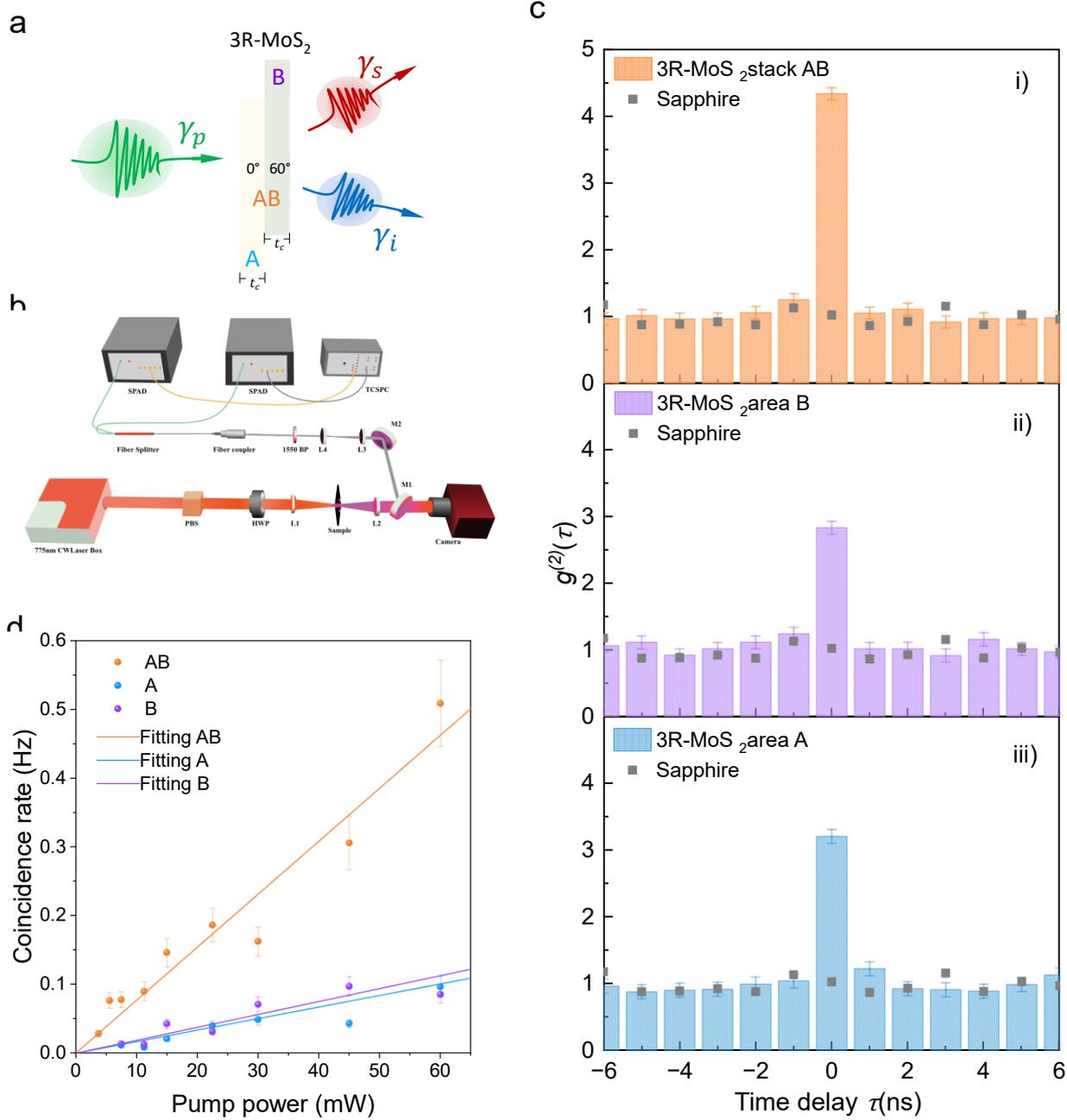

Figure 4

# FIGURE CAPTIONS

**Figure 1 | Modelling of Quasi-phase matching (QPM) by van der Waals stacking of 3R-MoS₂. a,** Schematic of second harmonic generation (SHG) through a 3R-MoS$_2$ flake, with a thickness of $t$. A fundamental pump with a wavelength at $\lambda$ is incident onto the flake, generating a second-harmonic (SH) signal with a wavelength at $\lambda/2$. **b,** Schematic of van der Waals stacking of two 3R-MoS$_2$ flakes, with a twist angle $\theta$. The top flake, bottom flake, and the stack area are labelled as B, A and AB, respectively. Each flake has a coherence length of $t_c$. **c,** Simulated SHG as a function of thickness $t$, for two scenarios: 1) non QPM (black line) and 2) QPM (pink line). For non QPM scenario, the SHG adds constructively (destructively) at a thickness of odd (even) times of the coherence length $t_c$. For the QPM scenario, multiple layers of 3R-MoS$_2$ (each layer with the same thickness of $t_c$) are stacked together. The orientation angles of the odd and even number of layers are set to be 0° and 60°, respectively.

**Figure 2 | QPM using two layers of 3R-MoS₂ with tunable twist angle $\theta$ . a,** Part i: Schematic of the measurement setup. The sample is pumped by a femtosecond laser at wavelength of 1550 nm, which goes through a long-pass filter, a $\lambda/2$ waveplate and a near infrared (NIR) objective lens. The SHG signal is collected by an identical NIR objective lens, passing through a linear polarizer before being sent to a spectrometer. Part ii: Zoom in schematic of the two 3R-MoS$_2$ flakes in experiments. By rotating the $\lambda/2$ waveplate, the polarization of incident laser was initially set to be along the armchair direction of the A flake, whose position was fixed. The top flake B was rotated on top of bottom flake A, by keeping a consistent physical gap of approximately 9.72 μm between flakes A and B. **b,** Polar plot for the measured SHG from three sample areas, bottom flake A (blue dots), top flake B (purple dots) and overlapping area AB (orange dots), as a function of the twist angle $\theta$. The thickness values of flakes A and B are 576 and 553 nm, respectively, close to the coherence length $t_c$. The corresponding solid lines are the simulation curves. The simulation is based on the model in Supplementary Note 9, and air gap is estimated as ~9.72 μm.

**Figure 3 | Experimental demonstration of QPM enabled by van der Waals stacking of 3R-MoS₂. a,** Optical microscope image of a 3R-MoS$_2$ homo-structure, formed by stacking two 3R-MoS$_2$ flakes, with a twist angle of 60°. The blue (purple) filled area highlights a uniform area with a measured thickness of 576 nm (564 nm) from bottom (top) flake A (B). The orange

filled area highlights the homo-structure region AB with a measured thickness of 1140 nm. Inset: Measured height vs distance profile along the red dashed line. **b,** Measured (dots) and simulated (solid line) co-polarized SHG as a function of the polarization angle of incident laser, from three areas A (blue), B (purple) and AB (orange) of the 3R-MoS$_2$ homo-structures with twist angles of 60°. The homo-structure with 60° twist angle is shown in (a). In experiment, the polarization of the incident laser was initially set to be along the armchair direction of the flake and controlled by rotating the $\lambda/2$ waveplate. A linear polarizer was used to select the polarization component of the SH radiation parallel to the polarization of the pump beam. **c,** Measured (black dots) enhancement factor as a function of the stack number ($N$) on the stacked van der Waals homostructure and the simulated curve $f(N) = N^2$ (Navy). The error bars represent the measurement uncertainty (standard deviations) obtained from more than ten data takes.

**Figure 4 | Enhanced spontaneous parametric down-conversion (SPDC) via QPM of 3R-MoS$_2$ homo-structure. a,** Schematic of SPDC process on 3R-MoS$_2$ van der Waals stacking. A single photon $\gamma_p$ with higher energy (1.60 eV) undergoes a nonlinear interaction in 3R-MoS$_2$ homo-structure with the stacked angle 60° (respect to the preceding 3R-MoS$_2$ sample in Figure 3a) and each layer at thickness of $t_c$, generating the entangled photon pair with lower energy (0.8 eV), known as the signal photon $\gamma_s$ and idler photon $\gamma_i$. **b,** Schematic of custom-built setup for photon-pair correlation measurements (orange pump photons and purple SPDC signals): 775 nm continuous wave (CW) pump generated from a single-frequency fibre laser, polarized beamsplitter (PBS), half-wave plate (HWP) for pump power control, spherical lenses, bandpass filter at 1550 nm with a 12 nm full-width at half-maximum, fibre coupler, and 50:50 fibre beamsplitter, SPAD; single-photon avalanche diode, time-correlated single photon counting (TCSPC) system. **c,** Normalized second-order correlation functions $g^{(2)}(\tau)$ from 3R-MoS$_2$ stack (i), 3R-MoS$_2$ area B (ii), area A (iii), and sapphire substrate (grey square dots) measured at 15 mW for 10 minutes with 1ns binning time. **d,** Comparison of power-dependent coincidence rate of the A (blue), B (purple) and AB areas (orange) of the 3R-MoS$_2$ homo-structure (with respect to the preceding sample in Fig 3a) measured for 10 minutes. The extracted slopes are 0.00771, 0.00168, 0.00188 for AB, A and B respectively. The error bars represent the standard deviations of SPAD counts per each measurement.